\documentclass[11pt]{article}
\usepackage[left=2.5cm,right=2.5cm,top=2.5cm,bottom=2.5cm]{geometry}
\usepackage{latexsym, amsmath, amssymb, mathrsfs, epsfig, graphicx, color}

\graphicspath{{eps/}}

\newcommand{\On}{{\rm O}(n)}
\newcommand{\half}{\frac{1}{2}}

\newcommand{\Zb}{\mathbb{Z}}
\newcommand{\SLE}{\rm SLE}
\newcommand{\PL}{P_{\rm left}}
\newcommand{\Cb}{\mathbb{C}}
\newcommand{\Pb}{\mathbb{P}}

\title{Spin interfaces in the Ashkin-Teller model and SLE}
\author{
         Y.~Ikhlef$^1$ and M.~A.~Rajabpour$^2$\footnote{e-mail: rajabpour@sissa.it}\\ \\
$^1$ Section Math\'ematiques, Universit\'e de Gen\`eve \\
        2-4 rue du Li\`evre, CP 64, 1211 Gen\`eve 4, Switzerland\\
        $^2$ SISSA and INFN, Sezione di Trieste, via Bonomea 265, 34136 Trieste, Italy 
     }

\begin{document}

\maketitle
\begin{abstract}
  We investigate the scaling properties of the spin interfaces in the
  Ashkin-Teller model. These interfaces are a very simple instance
  of lattice curves coexisting with a fluctuating degree of freedom,
  which renders the analytical determination of their exponents very difficult.
  One of our main findings is the construction of boundary conditions
  which ensure that the interface still satisfies
  the Markov property in this case.
  Then, using a novel technique based on the transfer
  matrix, we compute numerically the left-passage probability,
  and our results confirm that the spin interface is described by
  an SLE in the scaling limit. Moreover, at a particular point of the critical line, we
  describe a mapping of Ashkin-Teller model onto an integrable 19-vertex model,
  which, in turn, relates to an integrable dilute Brauer model. 
\end{abstract}
\vspace{5mm}
\begin{small}
\end{small}

\section{Introduction}
\label{sec:intro}

The scaling limit of interfaces in statistical models has attracted
much attention since the early developments of
conformal field theory
 (CFT). Starting from a local, classical
spin model -- {\it e.g.} the $\On$, Potts, $\Zb_N$, XY models --
there are essentially two ways of defining extended interfaces.
One may consider the polygons in a graphical high-temperature
expansion, {\it or} the spin interfaces (or domain walls) 
separating regions with different spin values.

As far as exact solutions are concerned, the situation turns out to be
very different for these two kinds of interfaces. On one hand, in 
several models, including the $\On$ and Potts
models, the high-temperature polygons
are described by a Coulomb gas (CG)~\cite{CG-Nienhuis,CG-Saleur}, {\it i.e.}
in the scaling limit, they become
the level lines of a free compact boson CFT. Many geometric properties of these
polygons can then be computed exactly through their mapping
to correlation functions in this CFT. This yielded also solid conjectures (and, in some
cases, rigorous proofs~\cite{Smirnov}) about their convergence to
a Schramm-Loewner Evolution (SLE)
model. On the other hand, despite some recent progress~\cite{Gamsa,Potts-DW},
no consistent theory was identified as the scaling limit of spin
interfaces, even in well-studied models such as the Potts model (except in
the particular case of the Ising model), and very few scaling properties of
these interfaces are actually known analytically.

Another current research direction is the study of ``extended'' SLEs,
which are curve models determined by two random processes: the driving
process of the Loewner chain, and an additional process 
corresponding, in the associated statistical model, to local variables
which are {\it not} fixed locally by the presence of the curve.
It was argued~\cite{ext-SLE1,ext-SLE2} that the martingale conditions for
this pair of processes correspond to null-state equations in an extended CFT.
In particular, the presence of the additional process affects the relation
between the central charge $c$ and the SLE parameter $\kappa$.
An important point that needs to be considered in this context,
is that in the presence of additional variables,
the lattice interface no longer satisfies the {\it Markov property},
which is an essential ingredient of the SLE model.

The Ashkin-Teller (AT) model is a simple local spin model,
with a critical line of constant central charge $c=1$ and varying exponents.
We believe it presents the two features described above: first, although it
has been much studied with both lattice and CFT techniques, the scaling theory for
its spin interfaces is not identified, and second, these interfaces leave some
fluctuating variables, which makes them good candidates for extended SLEs.

This paper presents two advances in the study of spin interfaces,
on the particular example of the AT model.
First, we introduce some specific boundary conditions
(BCs) which ensure that the spin interface satisfies the Markov
  property, even in the presence of an additional variable.
We then verify numerically
that our interface relates to SLE, and for this purpose, we introduce a new
algorithm allowing the measurement of the left-passage
probability~\cite{Schramm}
in the transfer-matrix formalism. Our results also include an accurate
numerical determination of the fractal dimension $d_f$ of spin interfaces,
also by means of the transfer matrix, which supports a conjecture
for $d_f$ stated in~\cite{CLR}.
Second,
we identify a specific value on the critical
line of the AT model, where the temperature tends to zero, and
at this point, we describe an exact mapping of the AT model onto
an integrable 19-vertex model. Although this second result does not give direct access to
spin interface exponents, this represents an alternative way of solving exactly
the AT model, and may lead to additional exact results on this model.

\section{Spin interfaces in the Ashkin-Teller model}

The Ashkin-Teller model is a two-parameter model with a self-dual,
critical line along which
the central charge is constant $c=1$ and the critical exponents vary. Its local
correlation functions are described by the $\Zb_2$-orbifold of the compact boson CFT.
However, like for the Potts model, this CG description only yields results for a
specific class of interfaces, which does not include the spin DWs. A striking
fact illustrating this is the absence of an exact result (except at specific values of
the couplings) for the fractal dimension of spin DWs along the critical line!

To be more precise, let us summarize here the current knowledge on spin interfaces
in the AT model. On the square lattice, the AT model consists of two spin variables
$\sigma_j$ and $\tau_j$ at each site $j$, taking the values $\pm 1$, and with the
Boltzmann weight
\begin{equation} \label{eq:W1}
  W[\sigma,\tau] = \prod_{\langle ij \rangle} \exp \left[
    \beta (\sigma_i\sigma_j + \tau_i\tau_j)
    + \alpha \ \sigma_i\sigma_j \ \tau_i\tau_j
  \right] \,,
\end{equation}
where the product is on all pairs of neighbouring sites.
The critical line for the square lattice is given by the condition:
\begin{equation}
  \sinh 2\beta = e^{-2\alpha} \,,
\end{equation}
which we parametrize as
\begin{equation} \label{eq:sd-line}
  2 - {\rm \coth}\ 2\beta = 1-e^{4\alpha} = 2 \cos \pi g\,,
  \qquad \text{with $0 \leq g \leq 1$.}
\end{equation}
To exhibit the $\Zb_4$ invariance of the model, it is convenient
to introduce the complex variables $S_j:= \frac{\sigma_j - i \tau_j}{1-i}$,
satisfying $S_j^4=1$, so that one can write~\eqref{eq:W1} as
\begin{equation} \label{eq:W2}
  W[S] = \prod_{\langle ij \rangle} \exp \left\{
    \beta (S_i^* S_j + S_j^* S_i)
    + \frac{\alpha}{2} \ \left[(S_i^* S_j)^2 + (S_j^* S_i)^2 \right]
  \right\} \,.
\end{equation}
This Boltzmann weight is invariant under the global rotation $S_j \to e^{\frac{i\pi}{2}} S_j$
and reflection $S_j \to S_j^*$. A non-branching spin interface in the AT model is
specified by the partition into two sets of the four possible spin configurations
$(\sigma,\tau)$:
\begin{equation}
  1: (++)
  \qquad 2: (+-)
  \qquad 3: (--)
  \qquad 4: (-+) \,.
\end{equation}
For example, $(14|23)$ denotes the interface between sites with values $1$ or $4$, and
sites with values $2$ or $3$. The symmetries of~\eqref{eq:W2} only leave three possible
nonequivalent non-branching spin interfaces:
\begin{itemize}

\item The case $(13|24)$ corresponds to domain walls for the variable $\tau'=\sigma\tau$.
  The associated operator has constant conformal dimension $h=\frac{1}{4}$ throughout
  the critical line, and hence the fractal dimension is $\frac{3}{2}$. In a previous
  work \cite{IR}, we constructed a discrete holomorphic parafermion $\psi_s$ of spin $s$
  in the AT model, which allowed us to relate the $(13|24)$ interface
  to $\SLE(4,\rho,\rho)$, where the value of $\rho$ was derived from the knowledge of $s$.
  The problem of the $(13|24)$ interface with various boundary conditions was also addressed
  numerically in~\cite{PS2} and~\cite{XOR}, yielding a set of conjectured
  relations to $\SLE(4,\rho',\rho')$, and confirming the above results.

\item The case $(1|234)$ is not understood for general couplings, but for $g=3/4$
  (where the AT model coincides with the integrable Fateev-Zamolodchikov $\Zb_4$ spin model),
  the associated operator satisfies a null-state equation in the $\Zb_4$-parafermionic CFT,
  and is argued \cite{PS1} to have fractal dimension $\frac{17}{12}$.

\item The case $(12|34)$ corresponds to domain walls for the variable $\sigma$. Its fractal
  dimension is only known for a few values of the parameter $g$:
  \begin{center}
    \begin{tabular}{c|c|c}
      $g$ & univ. class & $d_f$ \\
      \hline
      $1/4$ & ${\rm O}(n=2)$ & ${3}/{2}$ \\
%      $1/3$ & zero-temperature AT & $\mathbf{17/12}$ \\
      $1/2$ & Ising & ${11}/{8}$ \\
      $3/4$ & $\Zb_4$-parafermions & ${17}/{12}$ \\
      $1$ & four-state Potts & ${3}/{2}$
    \end{tabular}
  \end{center}
%  The value in bold is determined in the present paper.
  The fractal dimension cannot be simply obtained from the CG analysis.
  However, based on the above exact values and Monte-Carlo simulations (in the
  region between Ising and Potts points), the following expression
  was proposed~\cite{CLR} for $d_f$:
  \begin{equation} \label{eq:conj}
    d_f = \frac{7}{8} + \frac{g}{2} + \frac{1}{8g} \,.
  \end{equation}

\end{itemize}

\section{Spin interface with the Markov property}

\subsection{Definition}

Let $\Omega$ be a domain of the square lattice, and $a$ and $b$ two lattice
sites on the boundary. We want to find boundary conditions (BCs) such that:
\begin{itemize}

\item Every spin configuration on $\Omega$ contains a $(12|34)$ spin interface
  $\gamma$ going from $a$ to $b$. We denote the sites of the path $\gamma$ as
  $(\gamma_0, \gamma_1, \dots, \gamma_\ell)$, where $\gamma_0=a$,
  $\gamma_\ell=b$, and $\ell$ is the length of $\gamma$.

\item The induced probability measure on $\gamma$ satisfies the Markov property.

\end{itemize}

The {\it Markov property} for a probability measure $\mu_{\Omega,a,b}$ on paths
means that, for any time $s$, the measure conditioned on $(\gamma_0, \dots, \gamma_s)$
is the same as the measure defined in $\Omega \backslash (\gamma_0, \dots, \gamma_s)$:
\begin{equation}
  \forall s>0 \,,
  \qquad
  \mu_{\Omega,a,b}( \cdot |\gamma_0, \dots, \gamma_s)
  = \mu_{\Omega \backslash (\gamma_0, \dots, \gamma_s), \gamma_s, b}(\cdot) \,.
\end{equation}
In other words, any piece of curve connected to the boundary should act
like the boundary itself, when viewed from bulk variables.

Note that the BCs  associated to a $(12|34)$ spin interface -- {\it i.e.}
fixing the boundary spins to be in $\{1,2\}$ on one side of the
boundary between $a$ and $b$ and $\{3,4\}$ on the other side -- usually do
not produce a Markov curve as soon as $\alpha \neq 0$. Indeed, conditioning
on $(\gamma_0, \dots, \gamma_s)$ correctly fixes the $\sigma$ spins on the boundary
of $\Omega \backslash (\gamma_0, \dots, \gamma_s)$, but the resulting measure 
on spins includes interaction factors $-\alpha \tau_i \tau_j$
for neighbouring edges $\langle ij \rangle$ which cross $(\gamma_0, \dots, \gamma_s)$,
and hence the conditioned measure on $(\gamma_s, \dots, \gamma_\ell)$ is {\it not}
equal to the measure induced by the AT model
on $\Omega \backslash (\gamma_0, \dots, \gamma_s)$. We thus
need to introduce different BCs.

\begin{figure}[ht]
  \begin{center}
    \includegraphics{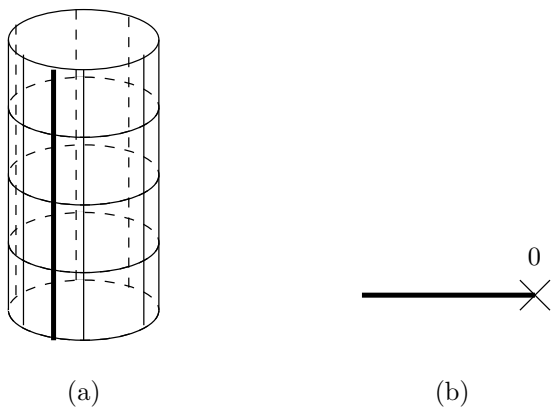}
    \caption{(a) Cylinder of circumference $L$ sites. The thick line
      is a cut, such that adjacent spins on its right have values in
      $\{1,2\}$, and adjacent spins on its left have values in
      $\{3,4\}$. The Boltzmann weights include interactions along edges which
      cross the cut, so that the spins live on the cylinder, while the
      $(12|34)$ interface lives on the strip. (b) The corresponding setting
      in the infinite plane: the domain for the interface is now the plane
      with a cut $\{z \in \Cb | z<0 \}$.}
    \label{fig:strip}
  \end{center}
\end{figure}

It is convenient to describe our BCs (see Fig.~\ref{fig:strip}a) on an infinite strip of width $L$ sites
$$S_L= \{ (x,y), x \in \{1, \dots, L\}, y \in \Zb \} \,.$$
Spins are fixed to be in $\{1,2\}$ on the left boundary, and
in $\{3,4\}$ on the right boundary. The Boltzmann weights are defined
as in~\eqref{eq:W1}, except that the product is now on
neighbouring pairs $\langle ij \rangle$ of the {\it cylinder} of width $L$,
{\it i.e.} it includes pairs of sites $\langle(1,y),(L,y)\rangle$.
In other words, the BCs are such that
the $\sigma$ spins live on a strip, and the $\tau$ spins on a cylinder.
This defines a Markov $(12|34)$ interface on the strip $S_L$, since
any piece of the interface connected to infinity acts exactly like a boundary.

\subsection{Scaling limit}

The above definition may be adapted to any domain $\Omega$ formed
by the complex plane with an infinite cut starting at the origin
(see Fig.~\ref{fig:strip}b):
the spins are then fixed to $\{1,2\}$ on one side of the cut,
and to $\{3,4\}$ on the other side, and we take the AT interaction
of the full plane.

In the scaling limit, the spin interface is expected to become
a continuous curve in $\Omega$, with a conformally invariant
distribution. Assuming that the Markov property is
preserved in this limit, Schramm's principle tells us that
the distribution of $\gamma$ is $\SLE_\kappa$, for some values of $\kappa$.
In the case $g=1/2$, $\gamma$ is an Ising spin interface, and its relation to
$\SLE_{3}$ is proved rigorously~\cite{Smirnov}.

Note that the above definition is restricted to a particular
type of domains, namely those where the boundary consists in a cut,
{\it i.e.} a possibly half-infinite simple curve. This is
because we want to include interactions across the boundary,
which automatically ensures the Markov property.
Extending this idea to a generic simply-connected domain
requires the introduction of a ``gluing'' map $f$ associated to,
$(\Omega, a, b)$, that is a continuous bijection between
the two arcs of the boundary defined by $a$ and $b$.
The measure should include an interaction between $z$ and $f(z)$
for all points $z$ on the boundary.
For instance, in the SLE process, if we denote $_t$ the guiding
process, $f_t$ is a one-to-one mapping between $\{z<a_t\}$ and
$\{z>a_t\}$, which is given by $f_0(z)=-z$ at the initial time $t=0$,
and evolves according to the Loewner process.
The presence of the process $f_t$ certainly affects the martingale
conditions for correlation functions, which in turn modifies the
relation between $\kappa$ and the central charge. In particular, in the AT model,
this accounts for the fact that the central charge
remains constant $c=1$, whereas $\kappa$ varies as a function of $g$.

More formally, one now needs to consider a measure $\mu_{\Omega,a,b}$ on
$(\gamma,f)$, and the Markov property reads
\begin{equation} \label{eq:Markov}
  \mu_{\Omega,a,b}(f, \cdot | \gamma_{[0,t]})
  = \mu_{\Omega \backslash \gamma_{[0,t]},\gamma_t,b}(\widetilde{f}^{(t)}, \cdot)
  \,,
  \qquad \text{where} \qquad
  \widetilde{f}^{(t)}(z) = \begin{cases}
    f(z) & \text{if $z \in \partial\Omega$} \\
    z & \text{if $z \in \gamma_{[0,t]}$.}
  \end{cases}
\end{equation}
Similarly, under a Loewner process $g_t$ in the half-plane, the function
$f_t$ evolves as
\begin{equation}
  f_t := g_t \circ \widetilde{f}^{(t)} \circ g_t^{-1} \,,
\end{equation}
where $\widetilde{f}^{(t)}$ is defined in~\eqref{eq:Markov}.
Assuming that all the relevant information about fluctuating
variables (in this case, the spins $\tau$ close to the boundary)
is encoded in $f_t$, the Markov property (and the associated martingale
conditions) should determine a non-standard relation between $c$ and $\kappa$.

\subsection{Numerical results}

For generic $g$, we check numerically that the spin interface becomes an
SLE in the scaling limit, using some known
properties of $\SLE_\kappa$. First, the fractal dimension of $\SLE_\kappa$
is given by
\begin{equation}
  d_f = 1 + \frac{\kappa}{8} \,.
\end{equation}
Also, the probability that $\gamma$ passes to the left of a point $(x,y)$ in
the strip $S_L$ is
\begin{equation}\label{schramm}
  \PL(x) = \half
  - \frac{\Gamma(4/\kappa)}{\sqrt{\pi}
    \ \Gamma \left(\frac{8-\kappa}{2\kappa} \right)}
  \ {\rm \cot} \left(\frac{\pi x}{L}\right)
  \ _2F_1 \left[
  \half,\frac{4}{\kappa},\frac{3}{2}, -{\rm \cot}^2 \left(\frac{\pi x}{L}\right)
  \right] \,.
\end{equation}
In the lattice model, both $d_f$ and $\PL$ can be computed numerically
by transfer-matrix diagonalisation, yielding two independent determinations
of $\kappa$. The fractal dimension is obtained from the
two-leg watermelon conformal dimension $X_2$ as $d_f = 2-X_2$, where $X_2$
is given by the dominant eigenvalue of the transfer matrix on the
cylinder in the sector where two interfaces are forced to propagate along
the cylinder. Moreover, we observe a better convergence when the model lives
on the triangular lattice, so we chose this setting for the determination of $d_f$.
The probability $\PL$ is computed from the components of
the dominant eigenvector of the transfer matrix on the strip, as described
in the Appendix.

The results for $d_f$ are shown in Fig.~\ref{fig:x2.tri}, and are in very
good agreement with the conjecture~\eqref{eq:conj}, on a large portion
of the critical line. Close to $g=1$, marginal operators typically
arise, and cause very strong finite-size effects, which explains the
poorer convergence in this region. The validity of~\eqref{eq:conj} in this
region was already checked by Monte-Carlo simulations in~\cite{CLR}.

\begin{figure}[ht]
  \begin{center}
    \includegraphics{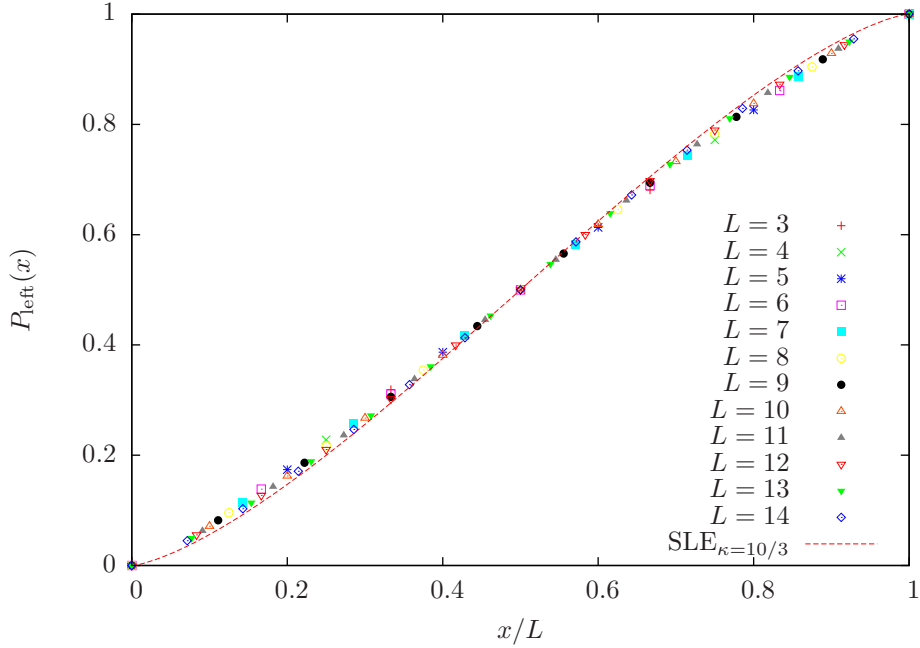}
    \caption{(Color online) Numerical determination of the left-passage probability~$\PL$ 
      in the AT model for $g=1/3$. The full line represents Schramm's formula,
      with the value $\kappa=10/3$ from the conjecture~\eqref{eq:conj}.}
    \label{fig:PL}
  \end{center}
\end{figure}

For the computation of $\PL$, we use the value $g=1/3$, where the
associated model is simpler, and hence the space of states on which
the transfer matrix acts has smaller dimension. The numerical results
on $\PL$, and the corresponding estimate of $\kappa$, are shown in
Fig.~\ref{fig:PL} and Table~\ref{table:kappa}. They match closely
with the results on $d_f$, which is a strong indication that the
curve distribution is $\SLE_{\kappa=10/3}$.

\begin{figure}[ht]
  \begin{center}
    \includegraphics{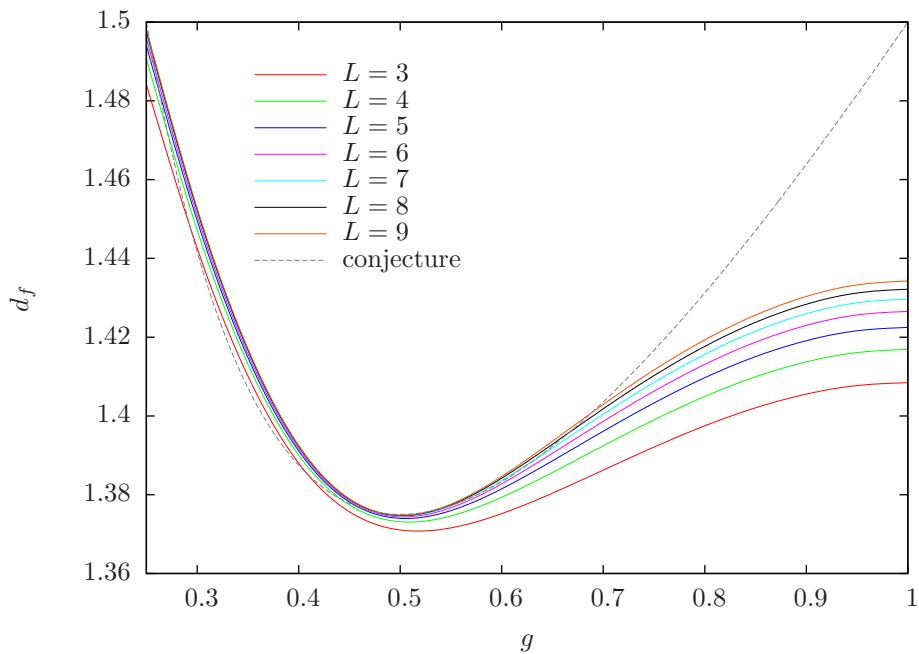}
    \caption{(Color online) Numerical determination of the fractal dimension $d_f$
      for the $(12|34)$ interface in the triangular-lattice AT model. The
      dotted curve represents the conjectured value~\eqref{eq:conj}.}
    \label{fig:x2.tri}
  \end{center}
\end{figure}
\begin{table}[ht]
  \begin{center}
    \begin{tabular}{l|l}
      $L$ &   $\kappa$ \\
      \hline
      3  &  3.74332 \\
      4  &  3.70126 \\
      5  &  3.66359 \\
      6  &  3.62997 \\
      7  &  3.59999 \\
      8  &  3.57317 \\
      9  &  3.54905 \\
      10 &  3.52724 \\
      11 &  3.50738 \\
      12 &  3.48919 \\
      13 &  3.47295 \\
      14 &  3.45693
    \end{tabular}
    \caption{Numerical estimates of $\kappa$ from the left-passage probability
      $\PL$ for $g=1/3$
      (see data in Fig.~\ref{fig:PL}). }
    \label{table:kappa}
  \end{center}
\end{table}

\section{The zero-temperature point}

\begin{figure}[ht]
  \begin{center}
    \includegraphics{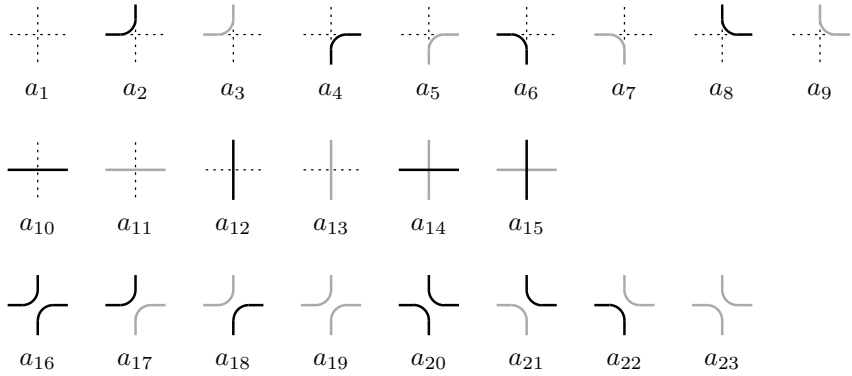}
    \caption{Vertices of the two-colour loop model.}
    \label{fig:loop}
  \end{center}
\end{figure}

The value $g=1/3$ on the self-dual line~\eqref{eq:sd-line} corresponds to the
limit
$$
\alpha \to -\infty, \qquad \beta \to \infty, \qquad
e^{\alpha+\beta}=\sqrt{2} \,.
$$
The Boltzmann weights then take the form
\begin{equation}
  e^{-(2\beta+\alpha) {\cal N}} \ W[\sigma,\tau] = \prod_{\langle ij \rangle}
  \left[
    \delta_{\sigma_i,\sigma_j} \ \delta_{\tau_i,\tau_j}
    + \half \delta_{\sigma_i,\sigma_j} \ (1-\delta_{\tau_i,\tau_j})
    + \half (1-\delta_{\sigma_i,\sigma_j}) \ \delta_{\tau_i,\tau_j}
    \right] \,,
\end{equation}
where $\cal N$ is the number of edges in $\Omega$.
The corresponding $\sigma$ and $\tau$ domain walls thus form a two-colour
loop model on the dual lattice, where a black (resp. gray) loop represents
a $\sigma$ (resp. $\tau$) domain wall. Each edge of the dual lattice
bears at most one colour of loop, and the weight of a loop configuration
is $K^{\ell}$, where $K=1/2$, and $\ell$ is the total length of the loops.
For completeness, we depict in Fig.~\ref{fig:loop} all the possible
loop configurations around a vertex.

\begin{figure}[ht]
  \begin{center}
    \includegraphics{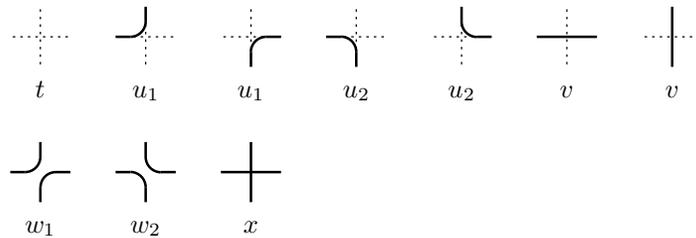}
    \caption{Vertices of the dilute Brauer model.}
    \label{fig:dilute-Brauer}
  \end{center}
\end{figure}

Let us now expose an exact equivalence of this two-colour loop model
of AT spin interfaces with an integrable 19-vertex (19V) model.
As a first stage, we map the 19V model onto a dilute Brauer model
with loop weight $n=2$. The vertices of the latter model are shown
in Fig.~\ref{fig:dilute-Brauer}. Each of the Brauer loops can be oriented
clockwise or anti-clockwise independently: this leads to a 19V model
with weights
\begin{eqnarray}
  \omega_1 &=& t \\
  \omega_2=\omega_3=\omega_4=\omega_5 &=& u_1 \\
  \omega_6=\omega_7=\omega_8=\omega_9 &=& u_2 \\
  \omega_{10}=\omega_{11}=\omega_{12}=\omega_{13} &=& v \\
  \omega_{14}=\omega_{15} &=& w_1+x \\
  \omega_{16}=\omega_{17} &=& w_2+x \\
  \omega_{18}=\omega_{19} &=& w_1+w_2 \,,
\end{eqnarray}
where we have used the standard notations $(\omega_1, \dots, \omega_{19})$ as in~\cite{IJMPB}.
If we take the FZ integrable weights \cite{IJMPB} of the 19V, we obtain the dilute Brauer
weights:
\begin{equation}
  \begin{array}{rcl}
    t &=& \sin \mu \sin 2\mu - \sin u \sin (\mu-u) \\
    u_1 &=& \sin 2\mu \sin (\mu-u) \\
    u_2 &=& \sin 2\mu \sin u \\
    v &=& \sin u \sin (\mu-u) \\
    w_1 &=& \sin 2\mu \sin (\mu-u) \cos u \\
    w_2 &=& \sin 2\mu \sin u \cos (\mu-u) \\
    x &=& -\cos 2\mu \sin u \sin (\mu-u) \,.
  \end{array}
\end{equation}
It turns out that setting $\mu=\frac{2\pi}{3}$ and $u=\mu/2$, and letting
$v \to -v$ (because $v$ configurations always appear in even number) yields
\begin{equation} \label{eq:Brauer-AT}
  \begin{array}{rcl}
    t &=& 1 \\
    u_1 = u_2 = v &=& K \\
    w_1 = w_2 &=& K^2 \\
    x &=& -K^2 \,.
  \end{array}
\end{equation}

The dilute Brauer model is, in turn, mapped to our two-colour loop model
in the following way. Each of the Brauer loops can be coloured black or gray
independently. Then, consider a vertex whose four adjacent edges bear the same
colour. Since the coloured loops have  weight unity, the overall Boltzmann
weight of the configuration is the same for vertices $w_1$,$w_2$ and $x$.
Thus the dilute Brauer model is equivalent to the two-colour loop model
provided the following relations hold
\begin{equation}
  \begin{array}{rcl}
    a_1 &=& t \\
    a_2= \dots= a_5 &=& u_1 \\
    a_6= \dots= a_9 &=& u_2 \\
    a_{10}= \dots= a_{13} &=& v \\
    a_{14} = a_{15} &=& \pm x \\
    a_{16} + a_{20} = w_{19} + w_{23} &=& w_1+w_2+x \\
    a_{17} = a_{18} &=& w_1 \\
    a_{21} = a_{22} &=& w_2
  \end{array}
\end{equation}
The integrable weights~\eqref{eq:Brauer-AT} satisfy these relations, and hence
we have proved the equivalence.

For the parameters $\mu=\frac{2\pi}{3}$ and $u=\mu/2$, the 19V model
is described~\cite{AM} in the continuum limit by a $c=1$ compact boson CFT
with the spectrum of conformal dimensions
\begin{equation}
  x_{n,m} = \frac{\mu}{2\pi} n^2 + \frac{\pi}{2\mu} m^2
  \,, \qquad n,m \in \Zb^2 \,.
\end{equation}
It turns out that the excitations described by this spectrum describes are not
directly related to those of the colored loops. In particular,
they do not include the conjectured two-leg exponent $x_2 = 7/12$.
However, the mapping to a 19V model may be exploited to obtain new exact
results on the AT model at zero temperature.

\section{Conclusion}

%In this paper we first found the fractal dimension of spin interfaces in the 
%Ashkin-Teller model in the region between Ising point and the
%XY point, the region missing in the work \cite{CLR}, by using transfer matrix
%technique. We confirmed that the fractal dimension of these interfaces follow
%closely the formula (\ref{eq:conj}). To show that these interfaces are
%actually conformally invariant we defined particular boundary conditions that
%have discrete domain Markov property. Then by inventing a novel technique
%based on the transfer matrix theory we were able to calculate the left-passage
%probability. Our numerical calculation shows that our interfaces follow the
%Schramm's formula (\ref{schramm}), in other words they are actually SLE.  Our
%technique of making interfaces with domain Markov property for models with
%internal symmetries is quite general and can be used for some other models,
%however, there is no guarantee for the preservation of this property in the
%continuum limit. 

In the present work, we have presented a set of new results on 
spin interfaces in the AT model.

By a numerical study based on transfer-matrix
diagonalisation, we have given a precise numerical determination of the
fractal dimension~$d_f$, in a parameter regime that could not be accessed
previously by a Monte-Carlo simulation~\cite{CLR}. Our results on $d_f$
bring more support to the conjectural expression~\eqref{eq:conj} introduced
in~\cite{CLR}. Moreover, we have shown, for the first time, how to adapt the
transfer-matrix method to measure numerically the left-passage probability
$\PL$, which is a natural quantity in the SLE literature.

On the analytical side, we introduce a general way of choosing BCs which
preserve the discrete Markov property of a path, even in the presence of additional
degrees of freedom which are not fixed locally by the path. This construction
is quite general, and may be extended to any lattice model with interfaces of
this sort.
The left-passage probability $\PL$ is then computed numerically with these
``Markovian'' BCs, and the results for $\PL$ confirm that the resulting
measure on curves becomes SLE in the continuum limit.

We have also identified a specific point on the critical
line of the AT model, the point where all the couplings diverge, 
where there is a mapping of the AT model onto
an integrable 19-vertex model. At this point the interfaces of the 
$\sigma$ and $\tau$ spins live in different bonds and so the corresponding 
two-color loop model has a simpler form.  At this particular point one can
also find a dilute Brauer model representation of the critical spin 
interfaces of the AT model. Since the loop model has a simpler form at 
this particular point, we believe it is a good place to look for a Coulomb
gas representation of the AT spin interfaces, and extend it to the
critical line afterwards.

Our work presents conclusive numerical arguments 
that the spin interfaces in AT model are conformally invariant and 
their fractal dimension follow the formula~\eqref{eq:conj},
motivating further the search for an analytic argument
which would account for~\eqref{eq:conj}.
This is a simple example of the more general notion of
spin interfaces, for which very few analytical results are available.
On top of the recent
methods for the study of lattice models, like 
discrete holomorphic parafermions, our construction
of ``Markovian'' BCs seems an interesting starting point for
this search.

\section*{Appendix: transfer-matrix algorithm for the left-passage probability}

In this Appendix, we present a numerical procedure for the computation
of the left-passage  probability in a loop model defined on the
infinite strip of width $L$ sites. Let us describe how our method works
on a simple loop model, such as the square-lattice $\On$ loop model (defined
by the vertices in Fig.~\ref{fig:dilute-Brauer}, with $x=0$). The
generalization to other loop models, such as the two-colour loop model studied
in this paper, is straightforward.

\begin{figure}[ht]
  \begin{center}
    \includegraphics{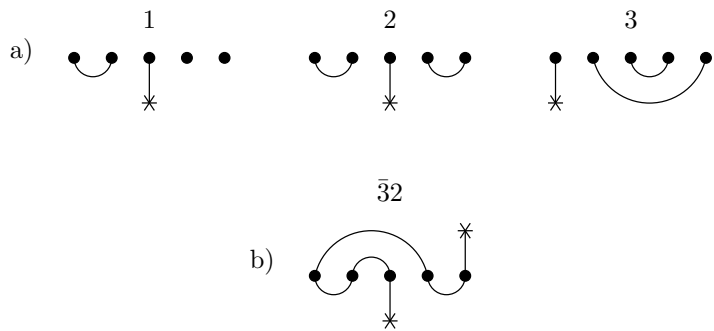}
    \caption{a) Some basis elements for $L=5$. b) An example of the gluing
      operation involved in the definition of $\langle \cdot, \cdot \rangle$.}
    \label{fig:states}
  \end{center}
\end{figure}

The basis elements $\alpha$ generating the space of states $V_L$ are
labelled by planar pairings of the points $\{1, \dots, L\}$, with one
point paired with $\star$, as depicted in
Fig.~\ref{fig:states}. The space $V_L$ is also equipped with a scalar product
$\langle \cdot, \cdot \rangle$, defined as follows. We say that two pairings
$\alpha,\beta$ are compatible if they have the same subset of paired points,
and we denote $\bar{\alpha}\beta$ the figure obtained by gluing the
$\pi$-rotation of $\alpha$ to $\beta$. Then
\begin{equation}
  \langle \alpha, \beta \rangle := \begin{cases}
    n^{\# {\rm closed \ loops \ in} \ \bar{\alpha}\beta}
    & \text{if $\alpha$ and $\beta$ are compatible,} \\
    0 & \text{otherwise.}
  \end{cases}
\end{equation}

Let us recall briefly how the transfer matrix $T$
acts on $V_L$~\cite{Blote-Nienhuis}.
We take the convention that $T$
acts from bottom to top, and adds one row to the system.
When starting from a pairing configuration $|\beta\rangle$, the 
addition of one row gives rise to a linear combination
of pairings $\sum_\alpha T_{\alpha \beta} |\alpha\rangle$,
where the matrix element $T_{\alpha \beta}$
is given by the product of Boltzmann weights inside the row,
including a factor $n$ for each loop that got closed in the process.

We denote the dominant eigenstate of $T$ in the sector with
a propagating path as
\begin{equation} \label{eq:psi}
  |\psi\rangle = \sum_\alpha \psi_\alpha |\alpha\rangle \,.
\end{equation}
If we cut the system into two
half-infinite strips above and below a given horizontal section,
then the $L$ dangling edges of the top half have a pairing $\bar{\alpha}$,
and those of the bottom half have a pairing $\beta$.
If we assume that the local Boltzmann weights in $T$ are invariant under
a rotation of angle $\pi$, then the probability of a pair
$(\bar{\alpha},\beta)$ is
\begin{equation}
  \Pb(\bar{\alpha},\beta) =
  \frac{\psi_\alpha \psi_\beta \langle \alpha, \beta \rangle}{Z} \,,
  \qquad \text{where} \qquad
  Z = \sum_{\alpha,\beta}
  \psi_\alpha \psi_\beta \langle \alpha, \beta \rangle
\end{equation}

For any pair $(\bar{\alpha},\beta)$, we denote
$\gamma_{\bar{\alpha},\beta}$ the open path appearing
in $\bar{\alpha}\beta$.
For a given position $j \in \{0, \dots, L \}$, we introduce the modified scalar
product $\langle \cdot, \cdot \rangle_j$:
\begin{equation}
  \langle \alpha, \beta \rangle_j := \begin{cases}
    n^{\# {\rm closed \ loops \ in} \ \bar{\alpha}\beta}
    & \text{if $\alpha$ and $\beta$ are compatible
      and $\gamma_{\bar{\alpha},\beta}$ passes between $0$ and $j$,} \\
    0 & \text{otherwise.}
  \end{cases}
\end{equation}
This allows us to express the left-passage probability $\PL$ as
\begin{equation} \label{eq:Pj}
  \PL(j) = \frac{1}{Z} \sum_{\alpha,\beta}
  \psi_\alpha \psi_\beta
  \ \langle \alpha, \beta \rangle_{j} \,.
\end{equation}

Our method consists in the following steps
\begin{enumerate}
  
\item Determine the eigenstate~\eqref{eq:psi} by the power method (iteration
  of $T$ on a random initial vector).

\item Compute~\eqref{eq:Pj} by performing the double sum on configurations.

\end{enumerate}

Since the computation of $\langle \alpha, \beta \rangle_{j}$ takes
${\rm O}(L)$ operation, the overall complexity of step 2 is
${\rm O}[L ({\rm dim} \ V_L)^2]$. 
The cost in memory is
${\rm O}({\rm dim} \ V_L)$, since only a finite number of vectors
are stored in memory, as usual in transfer-matrix algorithms.

\end{document}